\documentclass[useAMS,usenatbib]{mn2e}
\usepackage{url,times,graphicx,amsmath,amsfonts,amssymb,aas_macros,color,epsfig,epstopdf,bigints}
 \usepackage{xcolor}

\newcommand{\hMpc}{{\ifmmode{h^{-1}{\rm Mpc}}\else{$h^{-1}$Mpc}\fi}}
\newcommand{\hkpc}{{\ifmmode{h^{-1}{\rm kpc}}\else{$h^{-1}$kpc}\fi}}
\newcommand{\kpc}{{\ifmmode{ {\rm kpc} }\else{{\rm kpc}}\fi}}
\newcommand{\kms}{{\ifmmode{ {\rm km\,s^{-1}} }\else{ ${\rm km\,s^{-1}}$ }\fi}}
\newcommand{\hMsun}{{\ifmmode{h^{-1}{\rm {M_{\odot}}}}\else{$h^{-1}{\rm{M_{\odot}}}$}\fi}}
\newcommand{\Msun}{{\ifmmode{{\rm M}_{\odot}}\else{${\rm M}_{\odot}$}\fi}}
\newcommand{\Mhalo}{{\ifmmode{M_{\rm halo}}\else{$M_{\rm halo}$}\fi}}
\newcommand{\Rvir}{{\ifmmode{R_{\rm vir}}\else{$R_{\rm vir}$}\fi}}
\newcommand{\Mstar}{{\ifmmode{M_{\star}}\else{$M_{\star}$}\fi}}
\newcommand{\rs}{{\ifmmode{r_{\rm s}}\else{$r_{\rm s}$}\fi}}
\newcommand{\cSPH}{{\ifmmode{c_{\rm SPH}}\else{$c_{\rm SPH}$}\fi}}
\newcommand{\cDM}{{\ifmmode{c_{\rm DM}}\else{$c_{\rm DM}$}\fi}}
\newcommand{\rhos}{{\ifmmode{\rho_{\rm s}}\else{$\rho_{\rm s}$}\fi}}

\newcommand{\Vrot}{{\ifmmode{V_{\rm rot}}\else{$V_{\rm rot}$}\fi}}
\newcommand{\ltsima}{$\; \buildrel < \over \sim \;$}
\newcommand{\gtsima}{$\; \buildrel > \over \sim \;$}
\newcommand{\lsim}{\lower.5ex\hbox{\ltsima}}
\newcommand{\gsim}{\lower.5ex\hbox{\gtsima}}

\def\lesssim{\mathrel{\hbox{\rlap{\hbox{\lower4pt\hbox{$\sim$}}}\hbox{$<$}}}}
\def\gtrsim{\mathrel{\hbox{\rlap{\hbox{\lower4pt\hbox{$\sim$}}}\hbox{$>$}}}}

\newcommand{\Sec}[1]{Section~\ref{#1}}
\newcommand{\Eq}[1]{Eq.~(\ref{#1})}
\newcommand{\Fig}[1]{Fig.~\ref{#1}}
\newcommand{\beq}{\begin{equation}}
\newcommand{\eeq}{\end{equation}}

\def\beqa{\begin{eqnarray}}
\def\eeqa{\end{eqnarray}}

\def\head{ \vbox to 0pt{\vss \hbox to 0pt{\hskip 440pt\rm
      LA-UR-10-07069\hss} \vskip 25pt}}

\def\head{
 \vbox to 0pt{\vss
                   \hbox to 0pt{\hskip 440pt\rm LA-UR-10-07069\hss}
                  \vskip 25pt}}
\title[Mass-dependent dark matter halo profile]
{A mass-dependent density profile for dark matter haloes including the influence of galaxy formation}

\author[Di Cintio et. al]
{Arianna Di Cintio$^{1,2}$\thanks{E-mail: arianna.dicintio@uam.es}, Chris B. Brook$^1$, Aaron A. Dutton$^{3}$, Andrea V. Macci\`{o}$^3$, \newauthor Greg S. Stinson$^3$,  Alexander Knebe$^1$ \\
$^{1}$Departamento de F\'isica Te\'orica, M\'odulo C-15, Facultad de Ciencias, Universidad Aut\'onoma de Madrid, 28049 Cantoblanco, Madrid, Spain\\
$^2$Physics Department G. Marconi, Universit\`{a} di Roma Sapienza, Ple Aldo Moro 2, 00185 Rome, Italy\\
$^3$Max-Planck-Institut f\"ur Astronomie, K\"onigstuhl 17, 69117 Heidelberg, Germany\\
}

\begin{document}

\date{Accepted 2014 April 8.  Received 2014 April 8; in original form 2014 March 12}

\pagerange{\pageref{firstpage}--\pageref{lastpage}} \pubyear{2014}

\maketitle

\label{firstpage}


\begin{abstract}
  We introduce a mass dependent density profile to describe the
  distribution of dark matter within galaxies, which takes into
  account the stellar-to-halo mass dependence of the response of dark
  matter to baryonic processes.  The study is based on the analysis of
  hydrodynamically simulated galaxies from dwarf to Milky Way mass, drawn from the
  MaGICC project, which have been shown to match a wide range of disk
  scaling relationships.
  We find that the best fit parameters of a generic double power-law
  density profile vary in a systematic manner that depends on the
  stellar-to-halo mass ratio of each galaxy.  Thus, the quantity $\Mstar/\Mhalo$
  constrains the inner ($\gamma$) and outer ($\beta$) slopes of dark
  matter density, and the sharpness of transition between the slopes
  ($\alpha$), reducing the number of free parameters of the model to
  two. Due to the tight relation between stellar mass and halo
    mass, either of these quantities is sufficient to describe the dark matter
    halo profile including the effects of baryons. The concentration
  of the haloes in the hydrodynamical simulations is consistent with N-body
  expectations up to Milky Way mass galaxies, at which mass the haloes
  become twice as concentrated as compared with pure dark matter runs.
 
  This mass dependent density profile can be directly applied to rotation curve data of observed
  galaxies and to semi analytic galaxy formation models as a
  significant improvement over the commonly used NFW profile.
\end{abstract}

\noindent
\begin{keywords}

cosmology: dark matter galaxies: evolution - formation - hydrodynamics methods:N-body simulation 

 \end{keywords}

 \section{Introduction} \label{sec:introduction} 

Over several orders of magnitude in radius, dark matter (DM) halo density profiles arising from N-body simulations are well described by the so-called 'NFW' model  \citep{Navarro96,Springel08,Navarro10},  albeit with well known systematic deviations   \citep[e.g.,][]{Navarro04,Springel08, Gao08, Navarro10, DuttonMaccio2014}. The NFW function consists of two power laws, the inner region where the density is behaving  as $\rho\propto r^{-1}$ and the outer part as $\rho\propto r^{-3}$.
     
 The central $\rho\propto r^{-1}$ ``cusps'' of such model disagree with
 observations of real galaxies where mass modeling based on rotation curves finds much
 shallower inner density slopes, known as ``cored'' profiles
 \citep[e.g.,][]{moore94,Salucci00,deBlok01,Simon05,Deblok08,Kuzio08,Kuzio09,oh11b}. Cored galaxies are also found within the fainter, dark matter dominated dwarfs spheroidal galaxies surrounding the Milky Way \citep{Walker11}.
 This {\it cusp/core discrepancy} is usually seen as one of the
 major problems of the $\Lambda$CDM paradigm at small scales.

 The NFW profile is, however, derived from pure DM simulations in which particles only interact through gravity. 
  These simulations neglect hydrodynamical processes that may be
 relevant in determining the inner halo profile. Many
   studies have shown how baryons can affect the dark matter. Gas
 cooling to the center of a galaxy causes adiabatic contraction
 \citep[e.g.][]{Blumenthal86,Gnedin04}, whose effect
   strengthens cusps and exacerbates the mismatch between
    theoretical profiles and observations. Rather, expanded haloes are required to reconcile observed galaxy
 scaling relations of both early and late-type galaxies
 \citep{Dutton07,Dutton2013}.

 Baryons can expand haloes through two main mechanisms (see \citet{Pontzen14} for a recent review): outflows driven by stellar or AGN feedback
 \citep{Navarro96b,Mo04,Read05,Mashchenko06,Duffy10,Pontzen12,Martizzi13}
 and dynamical friction
 \citep{El-Zant01,Tonini06,Romano-Diaz08,DelPopolo09,DelPopolo10,Goerdt10,Cole11}.

 While dynamical friction is effective at expanding high mass
   haloes hosting galaxy clusters, stellar feedback is most effective
   at expanding low mass haloes \citep{governato10}.  Gas cools into
 the galaxy centre where it forms stars that drive repeated energetic outflows.
 Such outflows move enough gas mass to create a core in an originally cuspy dark halo, due to the DM response to the
 adjusted gravitational potential.  \citet{Penarrubia12} calculated the energy required to flatten a density profile as a function of halo mass.  The cusp/core change can be made permanent if the outflows are sufficiently rapid \citep{Pontzen12}.

Simulations from dwarf galaxies \citep{governato10,Zolotov12,teyssier13} to
  Milky Way mass \citep{Maccio12} have produced dark matter halo
  expansion depending on the implementation of stellar feedback.
\citet{Governato12} showed that only simulated galaxies with stellar
masses higher than $\sim 10^{7}M_\odot$ expand their haloes.
 They also showed that the inner DM profile slope, $\gamma$ in $\rho\propto r^{-\gamma}$, flattens with increasing stellar mass, resulting from the increase of available energy from supernovae.
An increase in stellar mass may, however, also deepen the potential well in the central region of the halo: indeed, \citet{DiCintio13} showed that above a certain halo mass such a deepened potential well opposes the flattening process.
  
\citet{DiCintio13} propose that $\gamma$ depends on the
stellar-to-halo mass ratio of galaxies. At
$\Mstar/\Mhalo\lesssim10^{-4}$ there is not enough supernova energy to efficiently change the DM distribution, and
 the halo retains the original NFW profile,
$\gamma\sim-1$. At higher $\Mstar/\Mhalo$, $\gamma$ increases, with the maximum $\gamma$ (most cored
galaxies) found when $\Mstar/\Mhalo$$\sim$$3-5\times10^{-3}$.  The
empirical relation between the stellar and halo mass of galaxies
\citep{moster10,guo10} implies that this corresponds to
$\Mstar$$\approx$$10^{8.5}M_\odot$ and
$\Mhalo$$\approx$$10^{11}M_\odot$.  In higher mass haloes, the outflow process becomes ineffective at flattening the inner DM density and the haloes have increasingly cuspy profiles.

In this paper, we take the next step to provide a mass-dependent parametrization of the entire dark matter density profile within galaxies.
 Using high resolution numerical simulations of galaxies, performed with the smoothed-particle hydrodynamics (SPH) technique, we are able to study the response of DM haloes to baryonic processes.  As with the
  central density slope $\gamma$ in \citet{DiCintio13}, we find that the density
  profile parameters depend on $\Mstar/\Mhalo$.

This study is based on a suite of hydrodynamically simulated galaxies, drawn from the Making
Galaxies In a Cosmological Context (MaGICC) project.  The galaxies
cover a broad mass range and include stellar feedback from supernovae,
stellar winds and the energy from young, massive stars.  The galaxies
that use the fiducial parameters from \citet{stinson13} match the
stellar-halo mass relation at $z=0$ \citep{moster10,guo10} and at higher redshift
  \citep{kannan13} as well as a range of present observed galaxy
properties and scaling relations \citep{brook12b,stinson13}. Unlike
previous generations of simulations, there is no catastrophic
overcooling, no loss of angular momentum \citep{brook11,brook12a}, and
the rotation curves do not have an inner peak, meaning that the mass
profiles are appropriate for comparing to real galaxies.

We present a profile that efficiently describes the distribution of
dark matter within the SPH simulated galaxies, from dwarfs to Milky Way
mass.  The profile is fully constrained by the integrated star
formation efficiency within each galaxy, $\Mstar/\Mhalo$, and the
  standard two additional free parameters, the scale radius $\rs$ and
the scale density $\rhos$ that depend on individual halo
  formation histories. After converting $\rs$ into $r_{-2}$, i.e. the point where the logarithmic slope of the profile equals $-2$, we
derive the concentration parameter for this new profile, defined as
$c=\Rvir/r_{-2}$, and show that for high mass galaxies  it substantially differs from expectation based on N-body
simulations.

This paper is organized as follows: the hydrodynamical simulations and feedback model are
presented in \Sec{sec:simulation}, the main results, including the derivation of profile parameters and galaxies rotation curves, together with a comparison with N-body simulations in \Sec{sec:results} and the conclusions in \Sec{sec:conclusions}.

\section{Simulations} \label{sec:simulation}
The SPH simulated galaxies we analyze here make up the Making Galaxies in a
Cosmological Context (MaGICC) project \citep{stinson13,brook12b}.
The initial conditions for the galaxies are taken from the
McMaster Unbiased Galaxy Simulations (\textsc{mugs}), which is
  described in \citealt{stinson10}.  Briefly, \textsc{mugs} is a
sample of 16 zoomed-in regions where $\sim$L$^\star$ galaxies form in
a cosmological volume 68 Mpc on a side.  \textsc{mugs} used a
$\Lambda$CDM cosmology with $H_0$= 73 \kms Mpc$^{-1}$,
$\Omega_{\rm{m}}=0.24$, $\Omega_{\Lambda}=0.76$,
$\Omega_{\rm{bary}}=0.04$ and $\sigma_8 = 0.76$
\citep[WMAP3,][]{Spergel07}. Each hydrodynamical simulation has a
corresponding dark matter-only simulation.

The hydrodynamical simulations used \textsc{gasoline}
\citep{wadsley04}, a fully parallel, gravitational N-body+SPH code. Cooling via hydrogen, helium, and
various metal-lines in a uniform ultraviolet ionizing background is
included as described in \citet{shen10}.

Standard formulations of SPH are known to suffer from some weaknesses \citep{Agertz07}, such as condensation of cold blobs which becomes particularly prominent in galaxies of virial masses $\sim 10^{12}M\sun$. 
We thus checked our results using a new version of \textsc{gasoline} which has a significantly different solver of hydrodynamics than the previous one (Keller et al. in prep). Within two simulated galaxies, which represent extreme cases (the cored most case and the highest mass case), we find that the dark matter density profiles are essentially identical to the ones found with the standard version of \textsc{gasoline}. As this new hydrodynamical code is not yet published, we have not included any figures here, but these preliminary tests give us confidence that our results are not predicated on the specific of the hydrodynamics solver. Indeed, it has been shown already that similar expansion processes are observed in galaxies simulated with grid-based codes \citep{teyssier13}.

The galaxies properties are summarized in Table 1: the sample comprises ten galaxies with five different initial conditions, spanning a wide range in halo mass. The initial conditions of the
medium and low mass galaxies are scaled down variants of the high mass
ones, so that rather than residing in a 68\,Mpc cube, they lie within
a cube with 34\,Mpc sides (medium) or 17\,Mpc sides (low mass). This rescaling allows us to compare galaxies with exactly the same merger
histories at three different masses. Differences in the underlying
power spectrum that result from this rescaling are minor
\citep{Springel08,maccio08,kannan12}. This assures us than any result derived from such sample, and presented in \Sec{sec:results}, will not be driven by the specific merger history.
It would be desirable, of course, to have a larger statistical sample of simulated galaxies and initial conditions, an issue that we hope to address in the near future.

The main haloes in our simulations were identified using the
MPI+OpenMP hybrid halo finder
\texttt{AHF}\footnote{http://popia.ft.uam.es/AMIGA}
\citep{Knollmann09,Gill04a}. \texttt{AHF} locates local over-densities
in an adaptively smoothed density field as prospective halo
centers. The virial masses of the haloes are defined as the masses
within a sphere containing $\Delta=92.8$ times the cosmic critical
matter density at $z=0$.

\subsection{Star Formation and Feedback}

\begin{table}
\begin{center}
  \caption{Properties of the SPH simulated galaxies used. $\Mhalo$ is the
    dark matter mass within the virial radius. The increasing symbol
    size indicates the membership of each galaxy to the low, medium or
    high mass group.}
\begin{tabular}{lllllllllllll}
\hline
 Mass   & ID  & soft  &$\Mhalo$               &$\Rvir$ &$\Mstar$           & sym \\
 range  &       & [pc]  &[M$_\odot$]             &[kpc]   & [M$_\odot$]       &  \\
\hline
 Low    & g1536 & 78.1  & $9.4$$\times$10$^{9}$  & 60 & $7.2$$\times$10$^{5}$ & \resizebox{.1cm}{.08cm}{$\color{black}\bullet$} \\
        &g15784 & 78.1 	& $1.9$$\times$10$^{10}$ & 77 & $8.9$$\times$10$^{6}$ & \resizebox{.1cm}{.08cm}{$\color{black}\blacktriangle $} \\
        &g15807 & 78.1 	& $3.0$$\times$10$^{10}$ & 89 & $1.6$$\times$10$^{7}$ & \resizebox{.1cm}{.08cm}{$\color{black} \blacksquare $}\\
\hline
 Medium & g7124 & 156.2 & $5.3$$\times$10$^{10}$ & 107 & $1.3$$\times$10$^{8}$ & $ \color{black}\ast $\\
        & g5664 & 156.2	& $6.3$$\times$10$^{10}$ & 114 & $2.4$$\times$10$^{8}$ & \resizebox{.15cm}{.1cm}{$\color{black}\blacklozenge $} \\
        & g1536 & 156.2 & $8.3$$\times$10$^{10}$ & 125 & $4.5$$\times$10$^{8}$ &  $\color{black} \bullet $ \\
        &g15784 & 156.2	& $1.8$$\times$10$^{11}$ & 161 & $4.3$$\times$10$^{9}$ &  $\color{black}\blacktriangle $ \\
\hline
High    & g7124 & 312.5 & $4.5$$\times$10$^{11}$ & 219 & $6.3$$\times$10$^{9}$  & \resizebox{.3cm}{.25cm}{$\color{black}\ast $}\\
        & g5664 & 312.5 & $5.6$$\times$10$^{11}$ & 236 & $2.7$$\times$10$^{10}$ & \resizebox{.25cm}{.2cm}{$\color{black} \blacklozenge $} \\
        & g1536 & 312.5 & $7.2$$\times$10$^{11}$ & 257 & $2.4$$\times$10$^{10}$ & \resizebox{.28cm}{.23cm}{$\color{black} \bullet $}  \\
\hline  
\end{tabular}\\
\end{center}
\label{tab:data}
\end{table}

The hydrodynamical simulations use the stochastic star formation recipe described in \citet{stinson06} in such a way that, on average, they reproduce the empirical Kennicut-Schmidt Law \citep{Schmidt59,kennicutt98}.

Gas is eligible to form stars when it reaches temperatures below T=15000 K and it is denser than 9.3 cm$^{-3}$, where the density threshold is set to the maximum density at which
gravitational instabilities can be resolved.

The stars feed energy back into
the interstellar medium (ISM) gas through blast-wave supernova
feedback \citep{stinson06} and ionizing feedback from massive stars
prior to their explosion as supernovae, referred to as ``early stellar
feedback'' \citep{stinson13}.

The implemented blastwave model for supernova feedback deposits $10^{51}$~erg into the surrounding ISM at
the end of the lifetime of stars more massive than 8\,M$_\odot$.
Since stars form from dense gas, this energy would be quickly radiated
away due to the efficient cooling. For this reason, cooling is delayed
for particles inside the blast region.  Metals are ejected from
Type~II supernovae (SNeII), Type~Ia supernovae (SNeIa), and the
stellar winds driven from asymptotic giant branch (AGB) stars, and
distributed to the nearest gas particles using the smoothing kernel
\citep{stinson06}. The metals can diffuse between gas particles as
described in \citep{shen10}.

Early stellar feedback is implemented using 10\% of the luminosity
emitted by massive stars prior to their explosion as supernovae. 

These photons do not couple efficiently with the surrounding ISM
\citep{freyer06}.  To mimic this inefficient energy coupling, we
inject $\epsilon_{\rm esf}$ of the energy as thermal energy in the
surrounding gas, and cooling is \emph{not} turned off, a procedure
that is highly inefficient at the spatial and temporal resolution of
cosmological simulations \citep{katz92,kay02}. Thus, the effective
coupling of the energy to the surrounding gas is only $\sim$ 1\%.

We analyze simulated galaxies that are part of the fiducial run of the
MaGICC project, which uses early stellar feedback with $\epsilon_{\rm
  esf}=0.1$ and a \citet{chabrier03} initial mass function. These simulations match the
abundance matching relation at $z=0$ \citep{moster10,guo10}, many present observed galaxy
properties \citep{brook12b,stinson13} as well as properties at high
redshift \citep{kannan13,Obreja14}.

\section{Results}\label{sec:results}

We analyze the dark matter density profiles of our SPH simulated
  galaxies using a five-free parameter $\alpha, \beta, \gamma$ profile
  function.  We show how to express $\alpha,
  \beta$ and $\gamma$ as functions of the integrated star formation
  efficiency $\Mstar/\Mhalo$ at z=0.

\subsection{$\alpha, \beta, \gamma$ profile}
The NFW profile is a specific form of the so-called
$(\alpha,\beta,\gamma)$ double power-law model \citep{Merritt06I,Hernquist90,Jaffe83}:

\beq
\rho(r)=\frac{\rho_s}{\left(\frac{r}{\rs}\right)^{\gamma}\left[1 + \left(\frac{r}{\rs}\right)^{\alpha}\right] ^{(\beta-\gamma)/\alpha}}
\label{five}
\eeq

\noindent
where $\rs$ is the scale radius and $\rho_s$ the scale density.
  $\rs$ and $\rho_s$ are characteristics of each halo, related to their mass and formation time
\citep[e.g.][]{Prada12,Munoz11,Maccio07,Bullock01}.  The inner and
outer regions have logarithmic slopes $-\gamma$ and $-\beta$,
respectively, while $\alpha$ regulates how sharp the transition
  is from the inner to the outer region. The NFW profile
  has $(\alpha,\beta,\gamma)=(1,3,1)$. In this case, the scale
radius equals the radius where the logarithmic slope of the density
profile is $-$2, $\rs = r_{-2}$. In the generic five-parameter model,

\beq
r_{-2}=  \left({\frac{2-\gamma}{\beta-2}}\right)^{1/\alpha}\rs\label{rho2rhos}
\eeq

\subsection{Constraining the halo profile via M$_{\star}$/M$_{halo}$ }\label{sec:constraining}
The dark matter halo profiles of each SPH simulated galaxy are computed in spherically averaged radial bins, logarithmically spaced in radius.
The number of bins $N_{\rm bin}$ in each halo is proportional to the number of particles within the virial radius, so that the best resolved haloes (with $\sim9$x$10^6$ particles) will have an higher $N_{\rm bin}$ with respect to the least resolved ones (with $3.5$x$10^5$ particles).

 We only considered bins within $0.01R_{\rm vir}<r<R_{\rm vir}$, as this region fulfills the
convergence criterion of \citet{Power03} in the least resolved
simulation. We perform a fitting procedure of the density profile using \Eq{five}, assigning errors to the density bins depending on the Poisson noise given by the number of particles within each shell, and using a Levenberg-Marquardt technique.

\begin{figure}
\hspace{-0.4cm}
\includegraphics[width=3.7in]{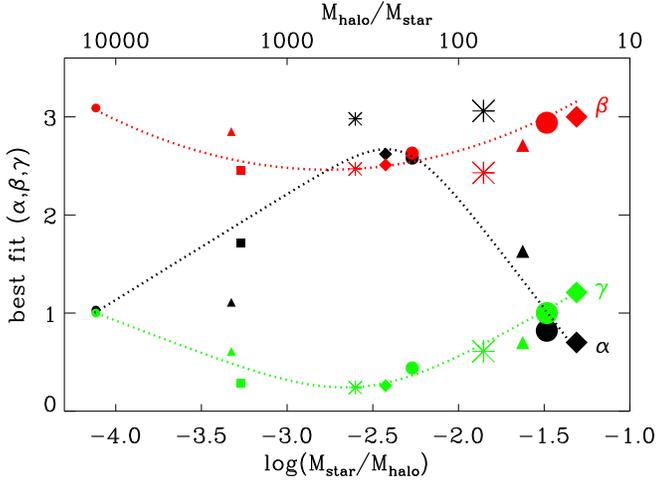}
\caption{Best fit parameters for the inner slope, $\gamma$
  (green), outer slope $\beta$ (red), and transition
  $\alpha$ (black) plotted as a function of integrated star formation
  efficiency, \Mstar/\Mhalo. The upper x-axis shows the corresponding \Mhalo/\Mstar\ as a reference to the mass to light ratio. The parameters are for the double power
  law model of the dark matter density profile in \Eq{five}.  Each SPH simulated galaxy is represented by a symbol of a different size and
  shape as described in Table 1.  The dotted lines represent the
  dependence of $\alpha$, $\beta$ and $\gamma$ on \Mstar/\Mhalo. 
  Their functional forms are given in \Eq{abg}.}
\label{fig:initial_param}
\end{figure}

\Fig{fig:initial_param} shows how the inner slope $\gamma$ (green),
the outer slope $\beta$ (red) and the transition parameter
$\alpha$ (black) vary as a function of the \Mstar/\Mhalo\ ratio.
The symbols, as explained in Table 1, correspond to different initial conditions, while their sizes indicate the mass of the halo. 
 The dotted lines show the best fit for each parameter, which we explain below in \Eq{abg}.

At very low integrated star formation efficiency, we expect to find
the same profile as a dark matter only simulation since star
  formation is too sporadic to flatten the profile. Indeed, at
$\log_{10}(\Mstar/\Mhalo)$$=$$-4.11$ the best fit values are
$\alpha$=1, $\beta$=3, and $\gamma$=1, exactly an NFW halo.

At higher integrated star formation efficiencies, both the inner
($\gamma$) and outer ($\beta$) profile slopes decline to lower
  values than an NFW model, indicating halo expansion. 
  At the same mass, the transition between inner and outer region becomes sharper: $\alpha$ increases as
high as 3. 
Thus, while baryonic processes affect the profiles mainly in the inner region of slope $\gamma$, we must take their effects into account when deriving the other  parameters $\alpha$ and $\beta$.

The star formation efficiency at which the cusp/core transition happens in our simulations is in agreement with the analytic calculation of \citet{Penarrubia12}, who compared the energy needed to remove a cusp with the energy liberated by SNeII explosions. 

The value of the inner slope ($\gamma$) varies with integrated
  star formation efficiency as found in \citet{DiCintio13}.  The
minimum inner slope is at $-2.6$$<$$\log_{10}(\Mstar/\Mhalo)$$<$$-2.4$. So,
as in \citet{DiCintio13}, the dark matter cusps are most efficiently
flattened when $\Mstar/\Mhalo\sim 3-5\times10^{-3}$.
Above $\log_{10}(\Mstar/\Mhalo)$$=$$-2.4$ (M/L$\sim$$250$),
 the parameters turn back towards the NFW values  since more mass
  collapses to the centre than the energy from gas can pull around.

We fit the correlation between $\alpha$, $\beta$, $\gamma$ and
the integrated star formation efficiency using two simple functions.
The outer slope, $\beta$, is fit with a parabola as a function of
  \Mstar/\Mhalo.  The inner slope, $\gamma$, and the transition
parameter, $\alpha$, are both fit using a double power law model as
a function of \Mstar/\Mhalo\ as in \citet{DiCintio13}. The best
  fit are shown as dotted lines in
Figure~\ref{fig:initial_param}. Their functional forms are:

\begin{equation}
\begin{aligned}
&\alpha= 2.94 - \log_{10}[(10^{X+2.33})^{-1.08}  +  (10^{X+2.33})^{2.29}]\\
&\beta=4.23+1.34X+0.26X^2\\
&\gamma= -0.06 + \log_{10}[(10^{X+2.56})^{-0.68}  +  (10^{X+2.56})]
\end{aligned}
\label{abg}
\end{equation}

\noindent where $X=\log_{10}(\Mstar/\Mhalo)$.

\Eq{abg} allows us to compute the entire dark matter profiles based
solely on the stellar-to-halo mass ratio of a galaxy.
We stress that the mass range of validity of \Eq{abg} is $-4.1<\log_{10}(\Mstar/\Mhalo)<-1.3$: at lower masses the ($\alpha,\beta,\gamma$) value returns to the usual (1,3,1), NFW prediction, while at masses higher than $10^{12}M_{\odot}$, i.e. the Milky Way, other effects such as AGN feedback can concur to modify the profile in a way not currently testable with our set of simulations.
In the future, having a larger statistical sample of simulated galaxies would certainly be desirable in order to compute the scatter in the relations defined by \Eq{abg}.

\subsection{Checking the $\alpha, \beta, \gamma$ constraints}
\begin{figure*}
\begin{center}
\vspace{-1.6cm}
\includegraphics[width=15cm,height=24cm]{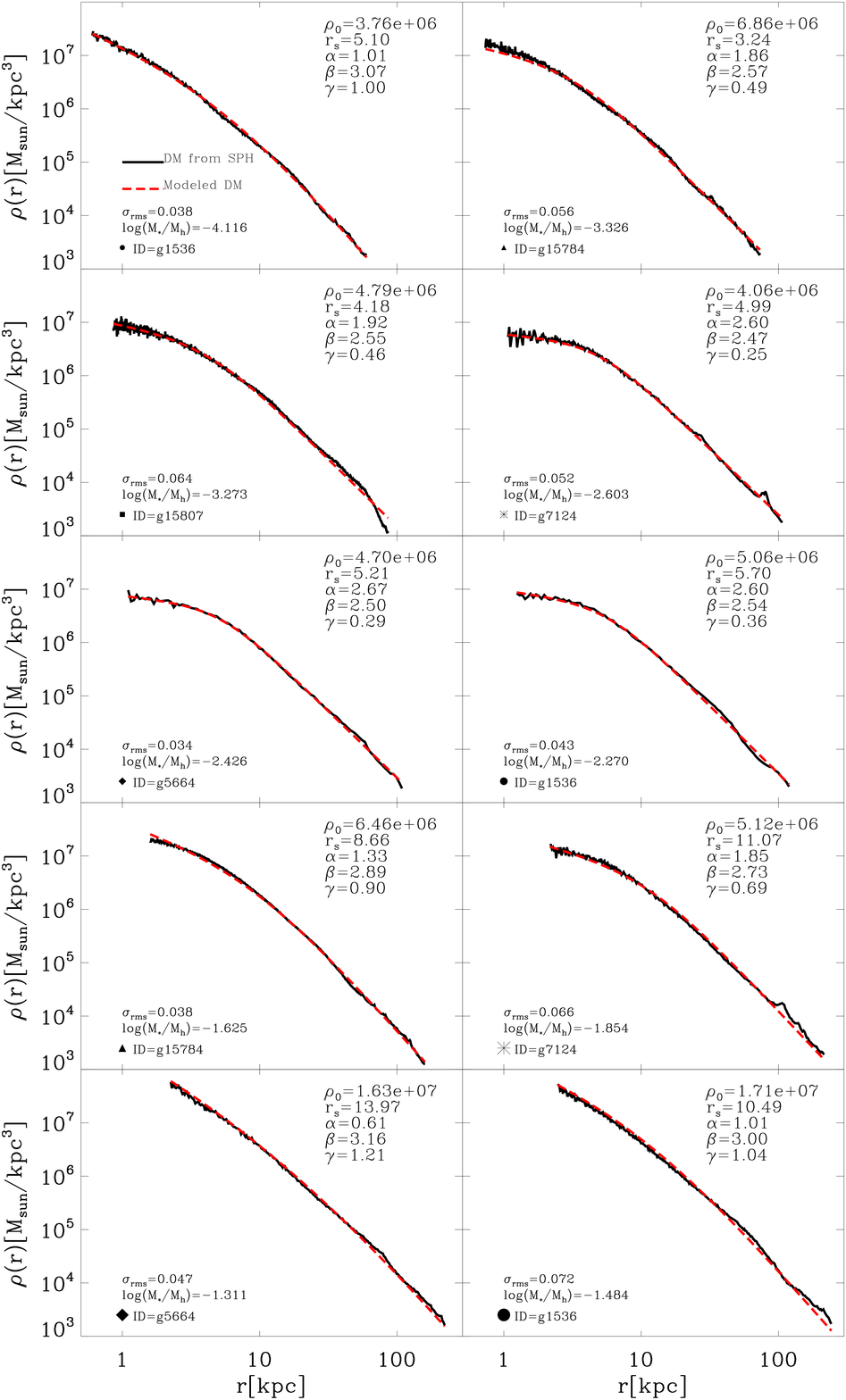}
\caption {Halo dark matter density profiles (black line) and
  best fit model (dashed red line) for the hydrodynamically simulated galaxies. The
  profiles start at $0.01\Rvir$ to ensure convergence and the galaxies
  are shown in increasing halo mass order, same as in Table 1. The
  constrained $\alpha$, $\beta$ and $\gamma$ values, from \Eq{abg}, are shown together
  with the corresponding efficiency \Mstar/\Mhalo. The two free
  parameters of the fit, $\rs$ and $\rhos$, are also listed as well as
  the r.m.s value of the fit $\sigma_{rms}$.}
\label{fig:density_fit}
\end{center}
\end{figure*}

 Using the constrained values for ($\alpha$, $\beta$, $\gamma$)
  from \Eq{abg}, we re-fit the dark matter density profiles of our
  haloes with the only standard two-free parameters, $\rs$ and
  $\rhos$. The fit results are shown as dashed red lines in
\Fig{fig:density_fit}, superimposed on the dark matter density
profiles of each hydrodynamically simulated galaxy (black lines). The
galaxies are ordered according to their mass from top left to bottom
right.  The best fit values obtained for the scale radius $\rs$ and
scale density $\rhos$ are shown in the upper-right corner, along with
the constrained values used for ($\alpha$, $\beta$, $\gamma$). The r.m.s. value of fit, defined as

\beq
\sigma_{rms}=\sqrt{\frac{1}{N_{\rm bins}}\sum_{k=1}^{N_{\rm bins}}(\rm log_{10}\rho_{sim,k} - log_{10}\rho_{fit,k})^2},
\eeq

are shown in the lower-left corner.
 The average value of $\sigma_{rms}$ is 0.051 and shows that \Eq{abg} can accurately describe the structure
  of simulated dark matter density profiles.

 Since we started our analysis using a five-free parameters model, it is possible that some degeneracies may exist,
and other combinations of ($\alpha,\beta,\gamma,\rs,\rho_s$) might be
equally precise in describing dark matter haloes. We do not claim that our model is unique, but rather that
provides a prescription that successfully describes very
different dark matter profiles, both cored and cusp ones, in galaxies.
 Our model, reduced to a two-free parameters profile using the value
 of \Mstar/\Mhalo\ (or simply \Mstar) of each galaxy, shows very good
 precision in reproducing halo density profiles of cosmological
 hydrodynamically simulated galaxies of any halo mass. 
 
\subsection{Modeling rotation curves}

 It is may be easier to compare observations with the dark matter rotation curves, rather than with the density profile.
   We proceed by deriving the quantity $V_{\rm c}(r)=\sqrt{GM(r)/r}$ for the dark matter component within hydrodynamical simulations, where 

\beq
M(r)=4\pi\rho_s\bigintss_0^r
\frac{r'^2}{\left(\frac{r'}{\rs}\right)^{\gamma}\left[1 + \left(\frac{r'}{\rs}\right)^{\alpha}\right] ^{(\beta-\gamma)/\alpha}}dr'
\eeq 

\noindent The values ($\alpha,\beta,\gamma$) are constrained through \Eq{abg} for each galaxy, while $\rho_s$ and $\rs$ are the best-fit results as listed in \Fig{fig:density_fit}, such that at the virial radius $M(\Rvir)$ equals \Mhalo.

\begin{figure*}\begin{center}
\vspace{-1.6cm}
\includegraphics[width=15cm,height=24cm]{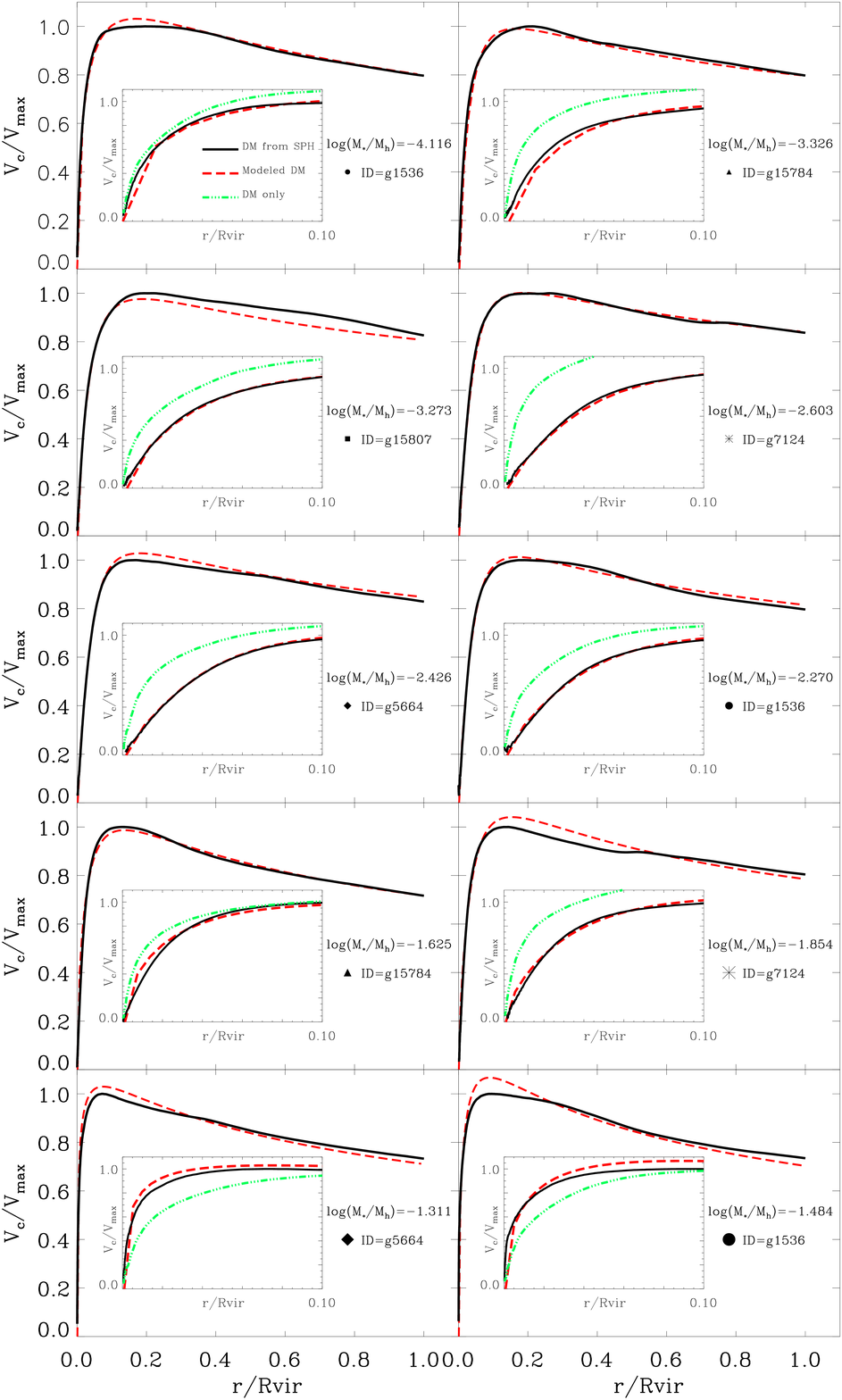}
\caption{Circular velocity curves of dark matter within the galaxies used in this
  work, $V_{\rm
    c}(r)=\sqrt{GM(r)/r}$.
 The dark matter rotation curve from the SPH run is shown as solid black line, while our parametrized model for describing it is shown as red dashed line.
 The small insert within each plot shows a zoom-in of the
  region within $0.1\Rvir$, with the addition of the rotation curve from dark
  matter only run as dotted-dashed green line. The $V_{\rm c}$ of each galaxy is
  normalized to its maximum values $V_{\rm max}$, and plotted in units
  of $\Rvir$. From left to right, and top to bottom, galaxies are ordered as in Table 1.}
\label{fig:velox_fit}
\end{center}\end{figure*}
    
The derived rotation curves for our model are shown as dashed red lines  in \Fig{fig:velox_fit}, with galaxies again ordered by mass as in  \Fig{fig:density_fit}.  The rotation curves taken directly from simulations, namely using the dark matter component within each hydrodynamically simulated galaxy, are shown as solid black lines. Each velocity curve is normalized to its maximum value $V_{max}$, and plotted in units of the virial radius.
              
The smaller panels within each plot show a zoom-in of $V_c(r)$ within $0.1\Rvir$, in order to better appreciate any difference between the actual simulations (solid black) and our parametrization (dashed red). Within this inner panel we also show as a green dotted-dashed line the rotation curve as derived from the dark matter only runs for each galaxy, scaled by the baryon fraction value. There is a very good agreement between our parametrized dark matter rotation curves and simulated ones, with differences that are below 10 per cent at any radii and for any galaxy. Further, when the contribution from the baryonic component is added to the rotation curves, the difference between the simulations and our parametrization will become even smaller, particularly at the high mass end of galaxy range where baryons dominate.   
By contrast, large differences can be seen between the rotation curves from dark matter only simulations (green dotted-dashed) and the rotation curves from the baryonic run (solid black) with the largest differences, as much as 50 per cent, being in intermediate mass galaxies.
Such differences highlighting the error one would commit by modeling rotation curves of real galaxies using prediction from N-body simulations, with a NFW profile unmodified by baryonic processes.
As opposite, our halo model introduces an error in the evaluation of galaxies' rotation curves which is well within observational errors, and can therefore safely be applied to model dark matter haloes within real galaxies.

\subsection{Constraining the concentration parameter}

 Now that we have demonstrated the precision of our density
profile based on the stellar-to-halo mass ratio as in \Eq{abg}, we
examine how one of the free parameters, the scale radius $\rs$, varies as
a function of integrated star forming efficiency, so that it could be
implemented in semi-analytic models of galaxy formation.  The
concentration parameter of our hydrodynamically simulated galaxies does
not always behave the same as in a corresponding dark matter only run.

 First, as $\alpha$, $\beta$ and $\gamma$ vary, the definition of
  $\rs$ changes.  For consistency, \Eq{rho2rhos} defines a conversion
  from $\rs$ to $r_{-2}$, the radius at which the logarithmic slope of
  the profile equals $-2$.  We define $\cSPH \equiv \Rvir/r_{-2}$ as
the concentration from the hydrodynamical simulation, and compare it
with $\cDM$, the NFW concentration from the dark matter only
simulation. 

\begin{figure}
\hspace{-0.4cm}
\includegraphics[width=3.6in,height=6cm]{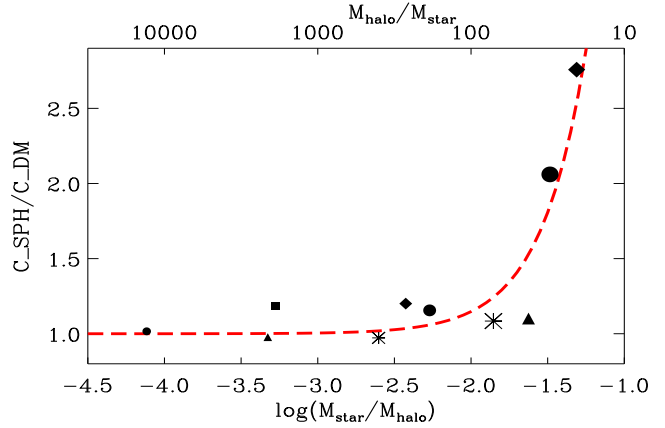}
\caption{Ratio between concentration parameter $c=\Rvir/r_{-2}$ in the
  SPH run and dark matter only run for our set of galaxies, as a
  function of \Mstar/\Mhalo. The upper x-axis shows the corresponding \Mhalo/\Mstar\ as a reference to the mass to light ratio. $\cDM$ has been derived fitting a NFW
  profile to the dark matter only version of each galaxy, while
  $\cSPH$ has been computed applying our model profile to the dark
  matter halo of the galaxies in the hydrodynamical run, and
  converting the corresponding $\rs$ into $r_{-2}$. The dashed red
  line represent the best model for the $\cSPH/\cDM$ values.}
\label{fig:conc_vs_magic}
\end{figure}

 \Fig{fig:conc_vs_magic} shows the ratio between the
  concentration parameter in the hydrodynamical simulation and the dark
  matter only one, and how this ratio varies as a function of \Mstar/\Mhalo. Each
simulation is represented by its symbol and size as described in Table
1.  The dependence of $\cSPH/\cDM$ on \Mstar/\Mhalo\ is nearly
exponential. The best fit is:
\beq
 \cSPH/\cDM=1.0 + 0.00003e^{3.4X}
 \label{concentration}
\eeq
where $X=\log_{10}(\Mstar/\Mhalo) + 4.5$.

Up to a mass ratio of $\log_{10}(\Mstar/\Mhalo) \sim -1.5$
(which corresponds to a halo mass of $10^{12} \Msun$),
$\cSPH$ is essentially the same as $\cDM$.  Thus, despite of the
variation of the inner slope, the transition to the outer slope happens at the same radius $r_{-2}$ as in the dark matter only simulation.

Above $\log_{10}(\Mstar/\Mhalo)\sim -1.5$, instead, the difference is striking and the haloes become much
  more concentrated in the SPH case than the corresponding DM only run.  In galaxies about the mass of the Milky Way, the
  inner region of the dark matter halo becomes smaller in our model, a
  signature of adiabatic contraction.
Indeed, as shown already in \citet{DiCintio13}, the increasing amount of
stars at the centre of high mass spirals opposes the flattening effect of
 gas outflows generating instead a profile which is increasingly
 cuspy and more concentrated.  
Collisionless simulations in a WMAP3 cosmology find that the
  typical concentration of a $10^{12}$ M$_\odot$ halo
  [$\log_{10}(\Mstar/\Mhalo)=-1.5$] is $c\approx8.5$ \citep{maccio08};  in our model with effective stellar feedback, the inner region of the halo shrinks by a factor of $\sim2$, giving a concentration parameter $\cSPH$ that can be $2.0-2.5$ times higher than the original N-body prediction. 
  
  Observations of the Milky Way are best fit with an NFW
halo with high concentration parameter $c\approx18-20$ \citep{Battaglia05,Catena10,Deason12a,Nesti13}.
 The data include halo tracers like globular clusters,
satellite galaxies, and dynamical observables like blue horizontal
branch stars, red giant stars and maser star forming regions used to constrain the Galactic potential. 
While such a high value of the concentration $c$ is at odds with respect to N-body predictions, our study suggests that the mismatch could be related to the effect of infalling baryons, and that a value of $c$ compatible with the above mentioned works it is indeed expected once such effect is properly taken into account in simulations.
Finally, a high concentration could arise possible tensions with the Tully-Fisher relation \citep{Dutton2011} and the Fundamental Plane \citep{Dutton2013} for high mass spirals, but this issue has to be explored in more detail once other effects relevant at L$^*$ scales, such as feedback from AGN, will be included in the simulations.

\section{Conclusions} \label{sec:conclusions}
It is well established that baryons affect dark matter density
profiles of haloes in galaxies \citep[e.g.][]{Blumenthal86,Navarro96b,El-Zant01,Gnedin04,Read05,Goerdt06,Read06,Mashchenko06,Tonini06,Romano-Diaz08,DelPopolo09,governato10,Goerdt10,DiCintio11,Zolotov12,Governato12,Maccio12,Martizzi13,teyssier13}.
  Simple arguments compare the energy available from star formation with the depth of a galactic potential
to estimate the degree of the change in the initial dark matter distribution \citep{Penarrubia12,Pontzen12,Pontzen14}.

This study describes the dark matter profiles of haloes from a suite of
hydrodynamical cosmological galaxy formation simulations that
  include the effects of stellar feedback. The profiles are modeled using a generic double
power law function. We find that the slope parameters of such model
  ($\alpha,\beta,\gamma$) vary in a systematic manner as a function
of the ratio between \Mstar/\Mhalo, which we call integrated star
formation efficiency. Using these fits allows us to propose a
star formation efficiency dependent density profile for dark matter
haloes that can be used for modeling observed galaxies and in
semi-analytic models of galaxy formation.
 
The star formation efficiency dependent density profile has the form
of a double power-law, with inner slope ($\gamma$), outer slope
($\beta$) and sharpness of transition ($\alpha$) fully determined by the
stellar to halo mass ratio as given in Eq.~\ref{abg}.  Thus, the five
free parameters of the generic model reduce to two, the scale
radius $\rs$ and scale density $\rhos$, the same free parameters of
the commonly used NFW model.

To examine how the scale radii varies as a function of integrated
  star formation efficiency, we compare the concentration parameter,
  $c=\Rvir/r_{-2}$, of the dark matter haloes from galaxies simulated
  with hydrodynamics prescriptions to those from the corresponding dark matter
  only simulations.
   For masses below roughly the Milky Way's the
  concentrations are similar, indicating that while the profiles may be significantly different from NFW, particularly in terms of inner slope, the radius at which the logarithmic slope of the profile equals -2 is the same as in the NFW model,
 indicating no net halo response at scales near the scale radius.

  However, for Milky Way mass galaxies
  the haloes from the hydro runs become as much as two times more
  concentrated than in the pure dark matter runs. 
  Such high concentrations are consistent to what has
  been derived from observations of Milky Way's dynamical tracers \citep{Battaglia05,Catena10,Deason12a,Nesti13}.
    
Thus, specifying the halo or stellar mass for a galaxy is
  sufficient to completely describe the shape of dark matter profiles
for galaxies ranging in mass from dwarfs to L$^*$, based on the influence of stellar feedback.
Importantly, the simulations we utilize in determining these profiles
match a wide range of scaling relations \citet{brook12b}, meaning that their radial
mass distributions are well constrained.

\noindent The main features of the mass dependent dark matter profile
are:
\begin{itemize}
\item Baryons affect the profile shape parameters.  For
  galaxies with flat inner profiles $\gamma$ the sharpness
  of transition parameter, $\alpha$, increases from 1 to 3 and corresponds to a small
  decrease in the slope of the outer profile $\beta$.
  \vspace{0.1cm}
\item At low integrated star formation efficiencies,
  $\Mstar/\Mhalo\lsim10^{-4}$ (galaxies with
  $\Mstar\lsim5$x$10^6\Msun$), dark matter haloes maintain the usual
  NFW profile as in dark matter only simulations.  \vspace{0.1cm}
\item At higher efficiencies the profile becomes progressively
  flatter.  The most cored galaxies are found at
  $\Mstar/\Mhalo\approx3-5\times10^{-3}$ or \Mstar\
  $\sim$$10^{8.5}$\Msun.  \vspace{0.1cm}
\item Galaxies with $\Mstar/\Mhalo\gtrsim5\times10^{-3}$
  ($\Mstar\gtrsim10^{8.5}\Msun$), become progressively steeper in the
  inner region as their mass increases.  \vspace{0.1cm}
\item The parameters ($\alpha,\beta,\gamma$) returns to the NFW values of
  (1,3,1) for L$^*$ galaxies.  \vspace{0.1cm}
\item However such L$^*$ galaxies, and more in general galaxies with $\Mstar/\Mhalo\gtrsim0.03$, are up to a factor of
  2.5 more concentrated than the corresponding dark matter only simulations.
\end{itemize}

In an Appendix we show step-by-step how to derive the dark matter profile for any galaxy mass.

Our results show that baryonic effects substantially change the structure of 
cold dark matter haloes from those predicted from dissipationless simulations, and therefore
must be taken into account in any model of galaxy formation.

Of course, our model uses a particular feedback implementation, namely thermal feedback in the form of blast-wave formalism. Yet \cite{teyssier13} finds a similar degree of core creation, at least in low mass galaxies, using a different feedback scheme. Both studies are based on the same mechanisms for core creation, i.e. rapid and repeated outflows of gas which result in changes in the potential. Indeed, the simulations closely follow the analytic model of core creation presented in \citet{Pontzen12}, indicating that the precise details of the feedback implementation are not central to our results, at least not in a qualitative manner.
 Galaxy formation models which do not include impulsive supernova explosions driving outflows from the central regions will not form cores in this manner.

In a forthcoming paper we will present a comprehensive comparison of our predicted density profile with the inferred mass distribution of observed  galaxies, with particular emphasis on Local Group members.

\section*{Acknowledgements}

We thank the referee for the fruitful report.
ADC thanks the MICINN (Spain) for the financial support through the MINECO grant AYA2012-31101.
She further thanks the MultiDark project, grant CSD2009-00064.
ADC and CBB thank the Max- Planck-Institut f\"{u}r Astronomie (MPIA) for its hospitality.
CBB is supported by the MICINN through the grant AYA2012-31101. 
CBB, AVM, GSS, and AAD acknowledge support from 
the  Sonderforschungsbereich SFB 881 ``The Milky Way System'' 
(subproject A1) of the German Research
Foundation (DFG).  
AK is supported by the {\it Ministerio de Econom\'ia y Competitividad} (MINECO) in Spain through grant AYA2012-31101 as well as the Consolider-Ingenio 2010 Programme of the {\it Spanish Ministerio de Ciencia e Innovaci\'on} (MICINN) under grant MultiDark CSD2009-00064. He also acknowledges support from the {\it Australian Research Council} (ARC) grants DP130100117 and DP140100198. He further thanks Gerhard Heinz for melodies in love.
We acknowledge the computational support provided by the UK's National
Cosmology Supercomputer (COSMOS), the {\sc theo} cluster of the
Max-Planck-Institut f\"{u}r Astronomie at the Rechenzentrum in
Garching and the University of Central Lancashire's High Performance
Computing Facility.  
We thank the DEISA consortium, co-funded through EU FP6 project
RI-031513 and the FP7 project RI-222919, for support within the DEISA
Extreme Computing Initiative.


\bibliographystyle{mn2e}
\bibliography{archive}

\bsp

\appendix
\section{Recipe to derive a mass dependent density profile}

We summarize here the steps necessary to derive, for a given halo mass, the corresponding dark matter profile which takes into account the effects of baryons:

\begin{itemize}
  \item Input the halo mass \Mhalo\ and the stellar mass \Mstar\ of a galaxy. In case that only one of these two quantities is known, use the abundance matching relation \citep{Brook14,Moster13,Guo11} to derive the second one.
  
  \item Specify an overdensity criterion, such that the halo mass is defined as the mass contained within a sphere of radius \Rvir\ containing $\Delta$ times the critical density of the Universe $\rho_{\rm crit}=3H^2/8\pi G$:

\beq
\Mhalo=\frac{4}{3}\pi R_{vir}^3\Delta\rho_{crit}
\eeq

Common choices of $\Delta$ are $\Delta_{\rm 200}=200$ or $\Delta_{\rm vir}=18\pi^2+82x-39x^2$ with $x=\Omega_m-1$ at $z=0$ \citep{Bryan98}. In a WMAP3 cosmology $\Delta_{\rm vir}=92.8$. 

\item Compute the halo profile parameters $(\alpha,\beta,\gamma)$ as a function of integrated star formation efficiency \Mstar/\Mhalo\
using \Eq{abg}. Recall that the range of validity of \Eq{abg} is $-4.1<\log_{10}(\Mstar/\Mhalo)<-1.3$: at lower efficiencies the ($\alpha,\beta,\gamma$) value returns to the usual (1,3,1), NFW prediction.

\item Obtain the concentration parameter $\cSPH=\Rvir/r_{-2}$ via \Eq{concentration}, where the quantity \cDM\ is the typical concentration of a halo of mass \Mhalo\ coming from N-body simulations \citep{DuttonMaccio2014,maccio08}. In this way we have derived the $r_{-2}$ at which the logarithmic slope of the profile equals -2. 

\item Convert such $r_{-2}$ into the corresponding scale radius \rs\, using \Eq{rho2rhos}. This is the scale radius that enters into \Eq{five}.

\item Find the scale density \rhos\ by imposing the normalization $M(<R_{vir})=\Mhalo$:

\beq
\rho_s=\Mhalo/ 4\pi \bigintss_0^{R_{vir}}
\frac{r^2}{\left(\frac{r}{\rs}\right)^{\gamma}\left[1 + \left(\frac{r}{\rs}\right)^{\alpha}\right] ^{(\beta-\gamma)/\alpha}}dr
\eeq

\item  The mass dependent density profile can now be obtained through \Eq{five} and the corresponding circular velocity via $V_{\rm c}(r)=\sqrt{GM(r)/r}$.

\item In case of fitting observed rotation curves of galaxies the scale radius \rs\ and scale density \rhos\ should be left as the two free parameters of the model.

\end{itemize}

\label{lastpage}

\end{document}